\documentclass[a4paper,11pt]{article}
\usepackage{pos}

\title{Effective LFV couplings at the electron-proton collider }

\author*[a]{Hrishabh Bharadwaj}
\author[a,b]{Sukanta Dutta}

\affiliation[a]{Government Mahila Degree College (Rajkiya Mahila Mahavidyalaya), Budaun-243601, U.P., India\\
Affiliated to MJP Rohilkhand University, Bareilly, U.P., India}

\affiliation[b]{SGTB Khalsa College, University of Delhi, Delhi-110007, India\\
Delhi School of Analytics, Institution of Eminence, University of Delhi,Delhi-110007, India}

\emailAdd{hrishabhphysics@gmail.com}
\emailAdd{sukanta.dutta@sgtbkhalsa.du.ac.in}

\abstract{We estimate the accuracy with which the coefficient of the lepton flavour-violating dimension-six operators can be measured at the proposed electron-proton collider.
Cuts-based analysis is performed to compute the signal significance at the centre of mass energy of 1.3 TeV, with a total integrated luminosity upto $\sim$ab$^{-1}$.  Using the optimal observables method for the kinematic distributions, we study the sensitivity of the effective couplings. We also study the impact of the initial electron beam polarisation.}

\FullConference{32nd International Workshop on Deep Inelastic Scattering and Related Subjects (DIS2025)\\
 24--28 March 2025\\
Cape Town, South Africa\\}

\begin{document}
\maketitle

\section{Introduction}

Lepton Flavour Violation (LFV) is one of the clearest indicators of physics beyond the Standard Model (SM). While LFV processes are highly suppressed in the SM due to tiny neutrino masses, many new physics scenarios predict observable rates in channels such as $\mu \to e\gamma$, $\tau \to \mu \mu \mu$, or lepton-flavour-violating meson decays \cite{Crivellin:2013hpa}. In this work, we explore the potential of a future electron-proton collider \cite{Dutta:2021del,Barik:2023bgx,Antusch:2020vul} to probe LFV interactions through effective field theory. We consider a set of dimension-six four-fermion operators that induce LFV transitions between electrons and muons. Using one-bin and optimal observables analyses, we estimate the sensitivity to the corresponding Wilson coefficients at $\sqrt{s} = 1.3\ \text{TeV}$ and an integrated luminosity upto $\sim \text{ab}^{-1}$. We also investigate the role of initial electron beam polarisation in improving these limits.

\section{Effective LFV Four-Fermion Operators and Phenomenology at the LHeC}
The SM is extended by dimension-six lepton flavour violating (LFV) four-fermion operators involving electrons, muons, and quarks. The effective Lagrangian is written as ${\cal L}_{\rm eff} = {\cal  L}_{\rm SM} + {\cal  L}_{\rm LFV}$,

\begin{eqnarray}
{\rm where}\ \ \ {\cal  L}_{\rm LFV}
&=& \frac{C^V_{LL}}{\Lambda^2} \left(\overline{e}\,\gamma^\alpha P_L\, \ell\right)\left(\overline{q}\,\gamma_\alpha P_L\, q\right)
+ \frac{C^V_{RR}}{\Lambda^2} \left(\overline{e}\,\gamma^\alpha P_R\, \ell\right)\left(\overline{q}\,\gamma_\alpha P_R\, q\right) \nonumber \\
&& +\,\frac{C^V_{LR}}{\Lambda^2} \left(\overline{e}\,\gamma^\alpha P_L\, \ell\right)\left(\overline{q}\,\gamma_\alpha P_R\, q\right)
+ \frac{C^V_{RL}}{\Lambda^2} \left(\overline{e}\,\gamma^\alpha P_R\, \ell\right)\left(\overline{q}\,\gamma_\alpha P_L\, q\right) \nonumber \\
&& +\,\frac{C^S_{LL}}{\Lambda^2} \left(\overline{e}\,P_L\, \ell\right)\left(\overline{q}\,P_L\, q\right)
+ \frac{C^S_{RR}}{\Lambda^2} \left(\overline{e}\,P_R\, \ell\right)\left(\overline{q}\,P_R\, q\right).
\end{eqnarray}
In this work, we focus on $\ell = \mu$. Here, $C_i$ are the Wilson coefficients and $\Lambda$ denotes the characteristic scale of new physics.
\par We study the LFV process
\begin{eqnarray}
e^- p \to \mu^- j,
\end{eqnarray}
which does not arise in the SM at tree level. For a benchmark setup with $C_i = 1$ and $\Lambda = 10$~TeV at $\sqrt{s} \simeq 1.3$~TeV, the signal cross sections for individual operators are listed in Table~\ref{tab:sig}.
\begin{table}[h!]\footnotesize
\centering
\begin{tabular}{|c|c|c|c|c|}
\hline
Chiral Operators & $\sigma^{C^V_{LL}}$ & $\sigma^{C^V_{RR}}$& $\sigma^{C^V_{RL}}$ & $\sigma^{C^S_{LL}}$\\
\hline
$\sigma_{\rm LHeC}$ (pb) &  $3.6\times 10^{-4}$ &$9.2\times 10^{-5}$  & $2.7\times 10^{-4}$ & $4.6\times 10^{-5}$ \\
\hline
\end{tabular}
\caption{\em{Signal cross sections at $\sqrt{s}\approx 1.3$~TeV. $C_i$s is set to unity  and cut-off scale $\Lambda$ is set to  $10$~TeV.}}
\label{tab:sig}
\end{table}
The event selection requires: $ p_T(j) \ge 5~{\rm GeV},\ \ p_T(\mu) \ge 2~{\rm GeV},\ \  
|\eta(j/\mu)| \le 4.5.$
The dominant SM backgrounds originate from leptonic decays of electroweak gauge bosons. Table~\ref{tab:bkgd} summarises the relevant contributions.

\begin{table}[h!]\centering\footnotesize
\begin{tabular}{|l|c|}
\hline
Background Process & $\sigma_{\rm LHeC}$ (pb) \\
\hline
$e^-p \to Z/\gamma^\ast(\to \mu^-\mu^+)\,\nu_\ell j$ & 0.029 \\
$e^-p \to W^\pm(\to \mu^\pm\nu_\mu)\,e^- j$ & 0.27 \\
$e^-p \to ZZ(\to \mu^-\mu^+)\,\nu_\ell j$ & $1.4\times 10^{-5}$ \\
$e^-p \to Z(\to \mu^-\mu^+)\,W^\pm(\to \mu^\pm\nu_\mu)\,\nu_\ell j$ & $2.5\times 10^{-5}$ \\
\hline
\end{tabular}
\caption{Dominant SM backgrounds to the LFV signal at the LHeC.}
\label{tab:bkgd}
\end{table}

\par
\begin{figure}
\includegraphics[scale=0.5]{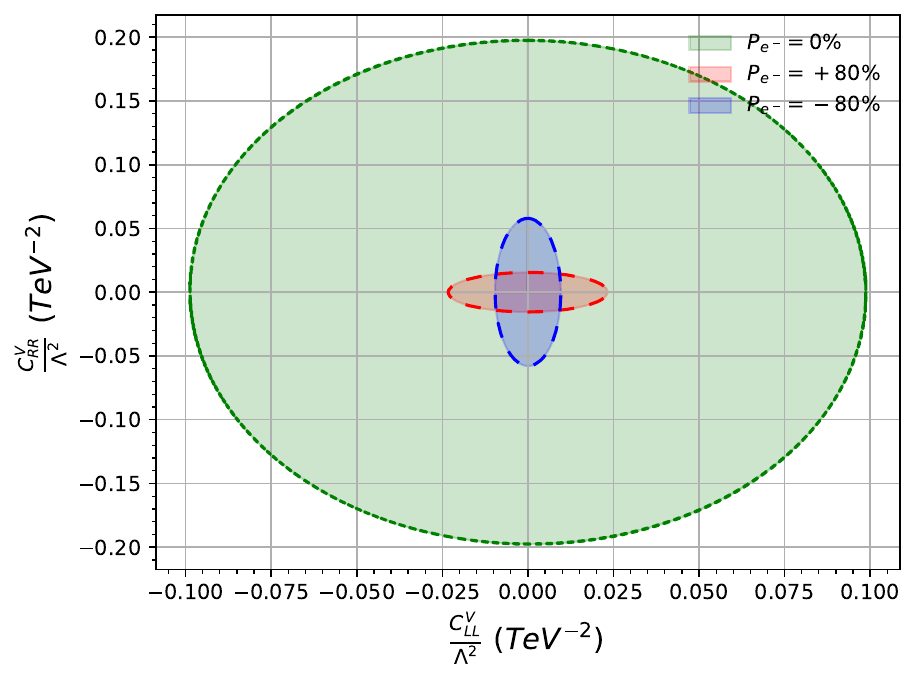}\includegraphics[scale=0.5]{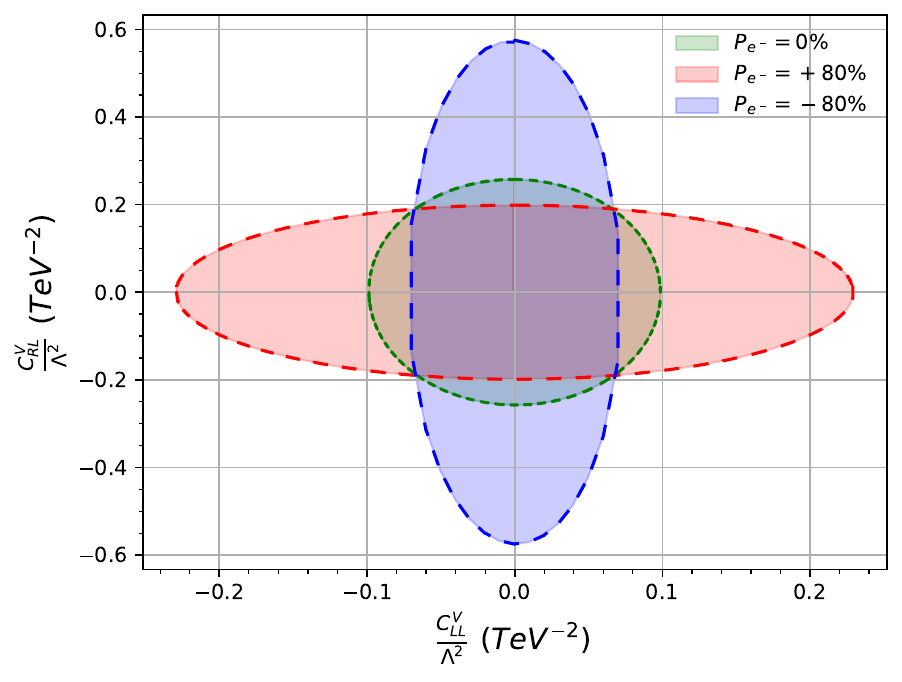}
\includegraphics[scale=0.5]{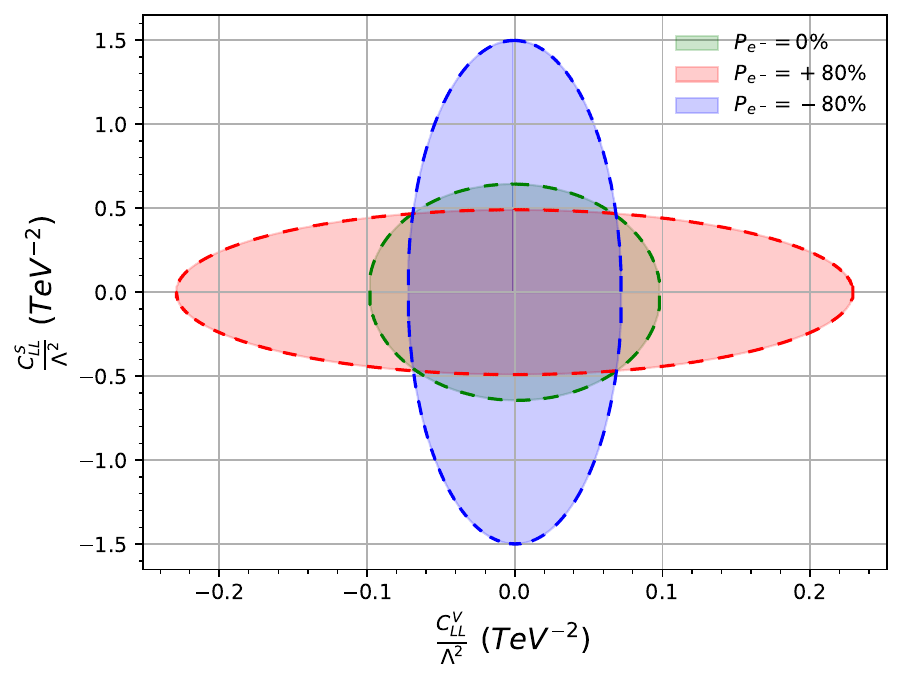}\includegraphics[scale=0.5]{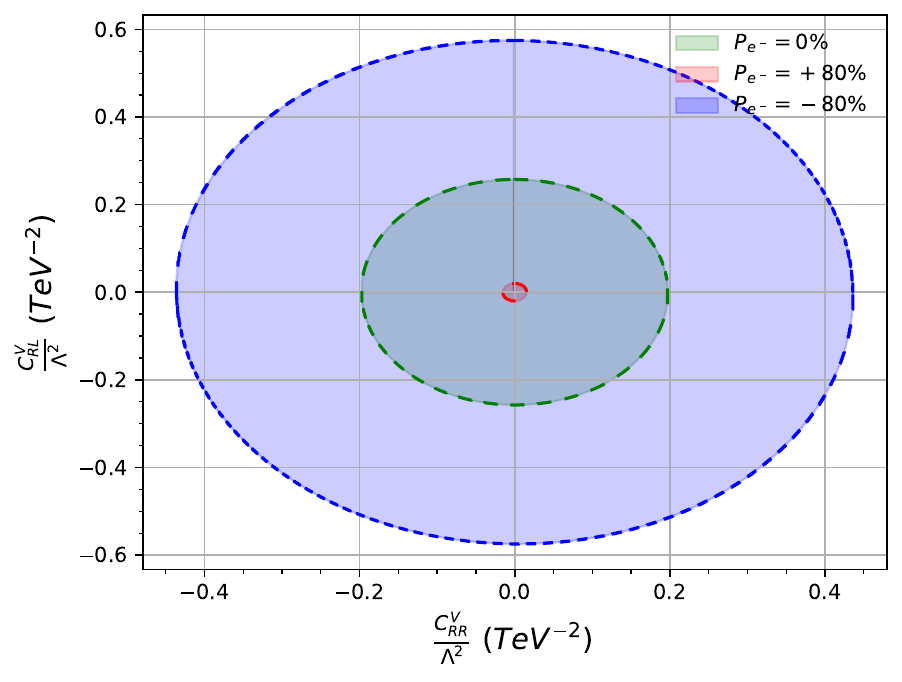}
\includegraphics[scale=0.5]{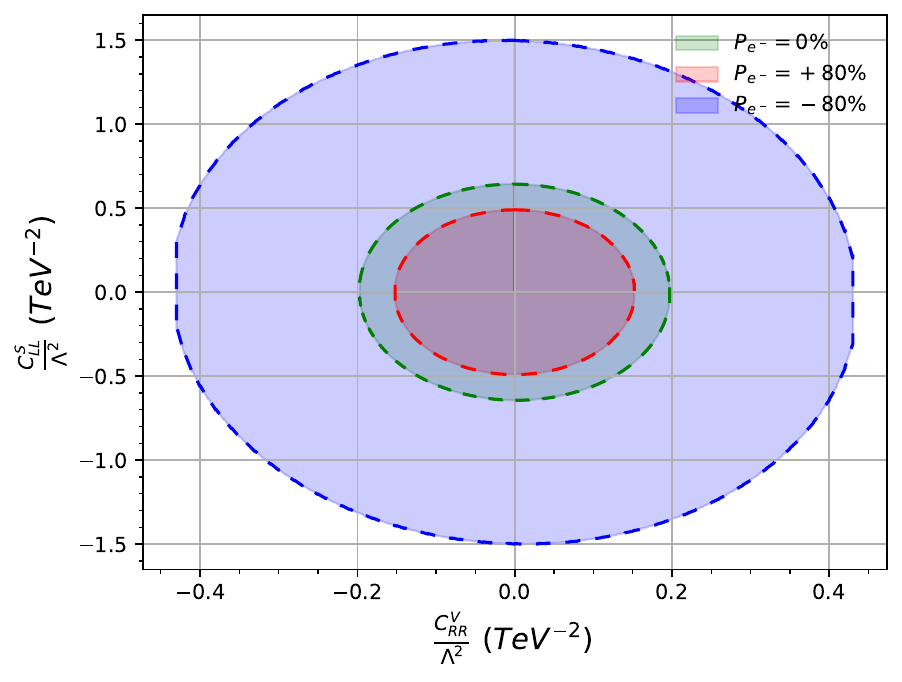}\includegraphics[scale=0.5]{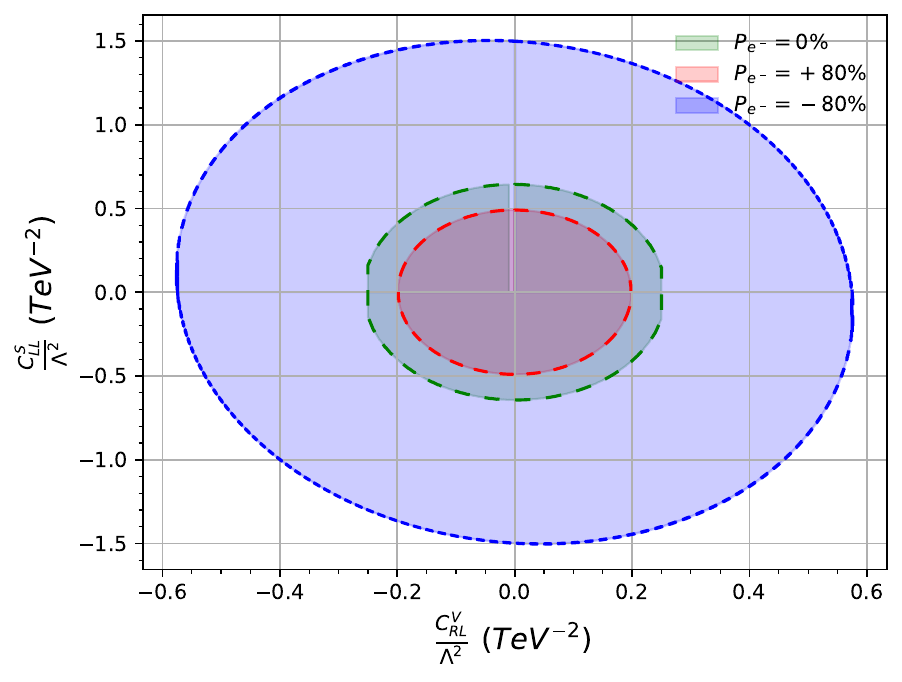}
\caption{5$\sigma$ discovery contours in the planes of different Wilson coefficient pairs.}
\label{5sigma}
\end{figure}

 We analyse the statistical significance of the Wilson's coefficients and are computed using computed as
\begin{eqnarray}
Z = \left[2\left((s+b)\ln\left(\frac{(s+b)(b+\Delta_b^2)}{b^2 + (s+b)\Delta_b^2}\right)
- \frac{b^2}{\Delta_b^2}\ln\left(1 + \frac{\Delta_b^2 s}{b(b+\Delta_b^2)}\right)\right)\right]^{1/2},
\end{eqnarray}
where $s$ and $b$ are the signal and background event yields, and $\Delta_b$ denotes the systematic uncertainty, taken to be $5\%$. The $5\sigma$ discovery contours in the planes of different Wilson-coefficient pairs are displayed in Fig. \ref{5sigma}.

\subsection{Optimal observable analysis}
\begin{figure}
\includegraphics[scale=0.5]{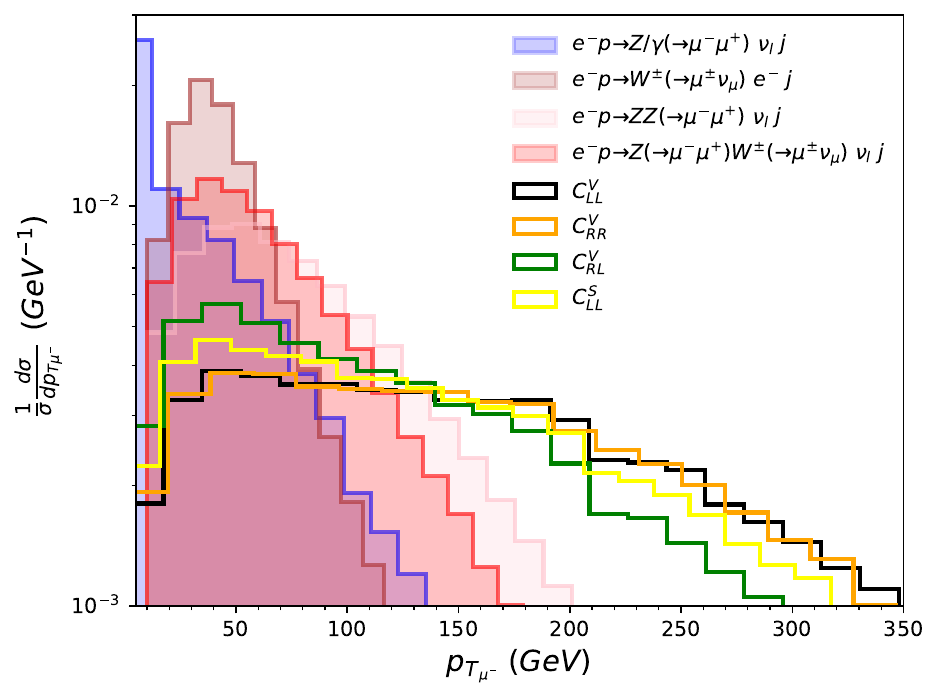}\includegraphics[scale=0.5]{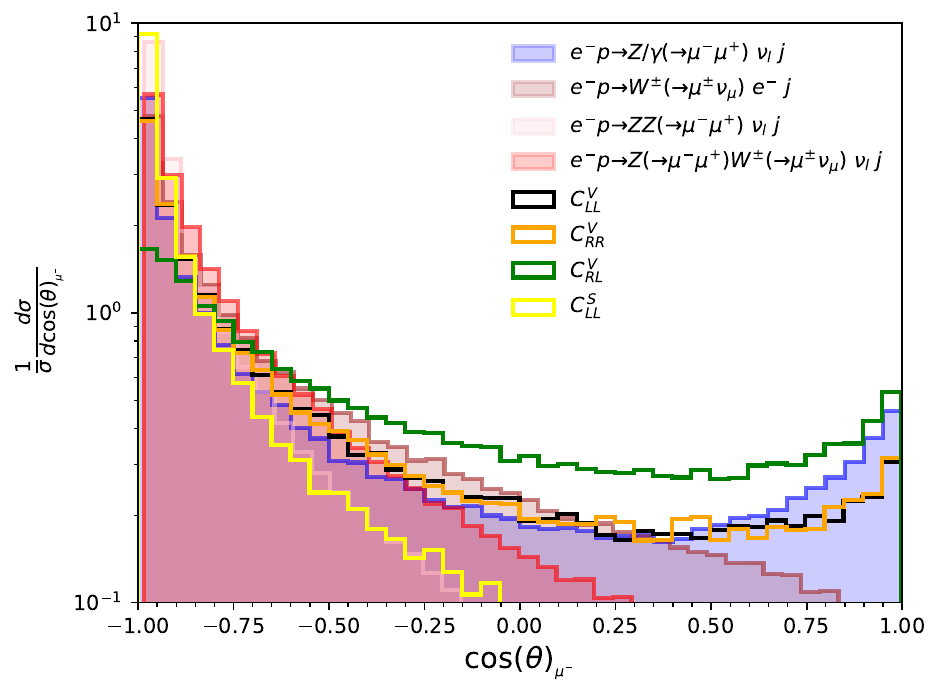}
\caption{Kinematic distributions used in the optimal observable analysis: transverse momentum $p_T$ (left) and scattering angle $\cos\theta$ (right) of the final state muon.}
\label{dbn}
\end{figure}
\par For any two couplings $c_i$ and $c_j$, the differential distribution of the events is given by:
      \begin{eqnarray}
      \frac{d^2N_{\rm total}}{dp_T\ d\cos\theta}&=& \frac{d^2N_{\rm SM}}{dp_T\ d\cos\theta}\ +\ \sum_i c_i^2\ \frac{d^2N_{c_i}}{dp_T\ d\cos\theta}\ + \sum_i \sum_{i\neq j}\ \frac{d^2N_{\rm c_i,c_j}}{dp_T\ d\cos\theta} 
      \end{eqnarray}
 For two couplings at a time, the covariance matrix is computed as:
      \begin{eqnarray}
      V_{ij}&=&F_{ij}^{-1} 
      \end{eqnarray}
      where
      \begin{eqnarray}
      F_{ij}&=& \sum_k^{\rm bins} \frac{1}{\Delta N_k^2} \int_k \left( \frac{\partial}{\partial c_i} \frac{d^2 N_k}{d p_T\ d\cos\theta} \right) \left( \frac{\partial}{\partial c_j} \frac{d^2 N_k}{d p_T\ d\cos\theta} \right) d p_T\ d \cos\theta,
      \end{eqnarray}
      $N_k$ denotes the total number of events in a bin.
 $\chi^2$ for the couplings $c_i$ and $c_j$ is given by:
      \begin{eqnarray}
      \Delta\chi^2 (c_i,c_j)&=& \sum_{i,j}\ \left(c_i^0 - c_i\right)\ V_{ij}^{-1}\  \left(c_j^0 - c_j\right)
      \end{eqnarray}

\begin{figure}
\includegraphics[scale=0.6]{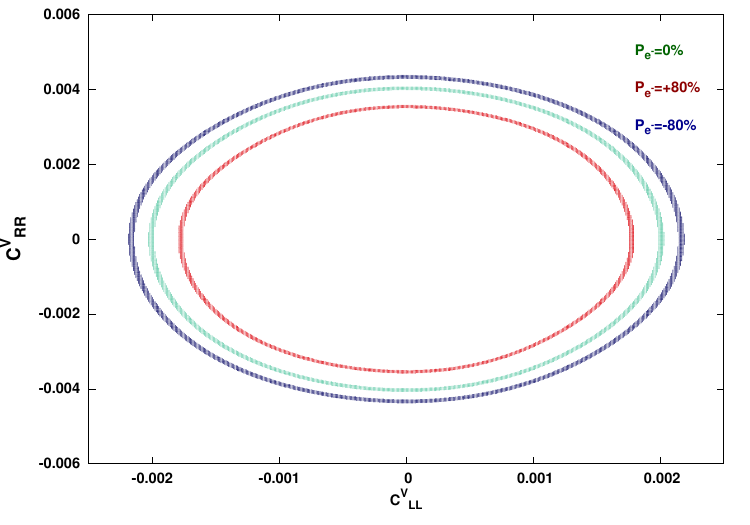}\includegraphics[scale=0.6]{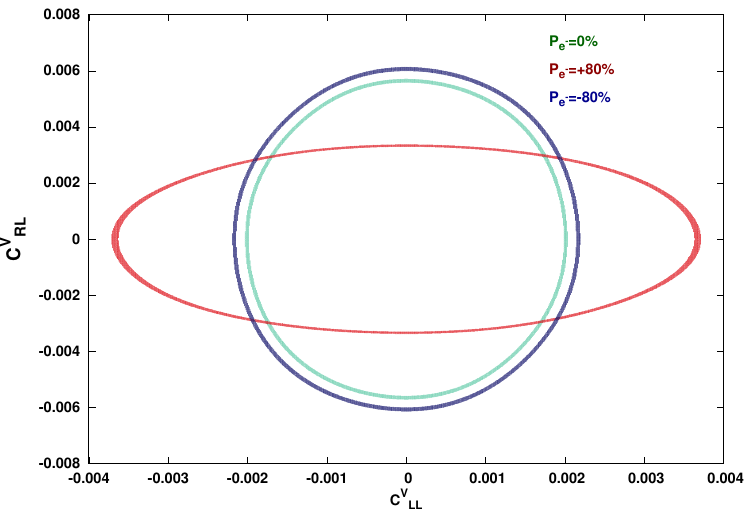}
\includegraphics[scale=0.6]{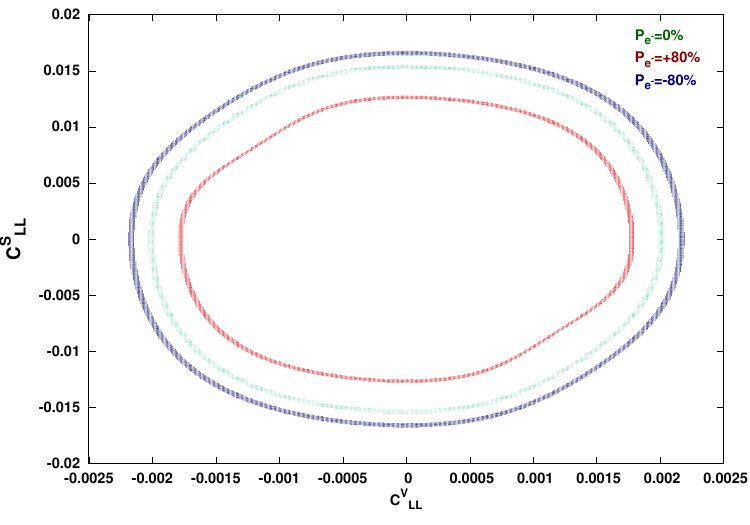}\includegraphics[scale=0.6]{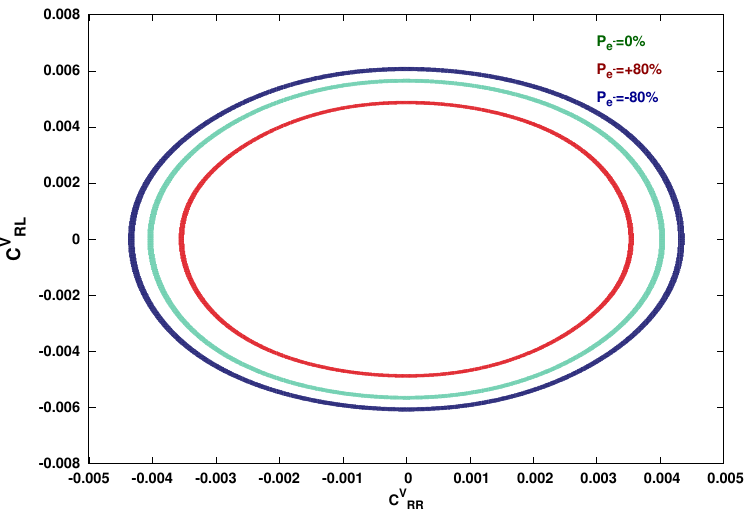}
\includegraphics[scale=0.6]{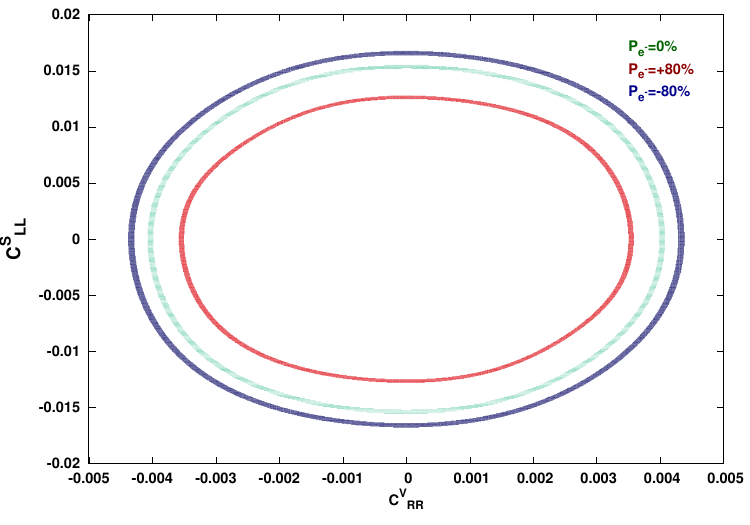}\includegraphics[scale=0.6]{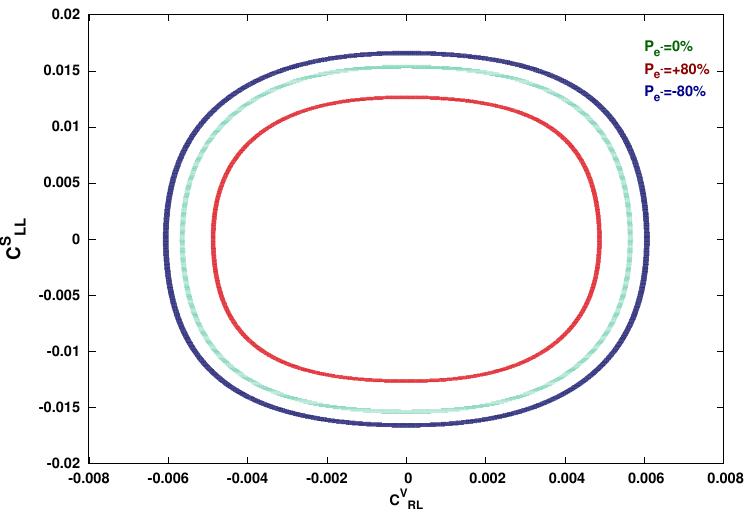}
\caption{$2\sigma$ exclusion contours obtained from the optimal observable analysis for different pairs of Wilson coefficients.}
\label{2Sigma}
\end{figure}

The optimal observable method \cite{Dutta:2008bh} exploits the full kinematic information of the final state, in particular the transverse momentum $p_T$ and scattering angle $\cos\theta$ of the muon, whose distributions are displayed in Fig. \ref{dbn}. We bin the differential distribution and construct a Fisher information matrix $F_{ij}$ from the derivatives of the expected event rates with respect to the Wilson coefficients. The inverse of this matrix gives the covariance matrix $V_{ij}$, which is used to define the $\chi^2$ function for parameter estimation. Fig. \ref{2Sigma} shows the $2\sigma$ contours in the planes of two operators at a time, marginalising over the others. The improvement in sensitivity compared to the one-bin analysis is evident, especially for operators with small cross sections.

\section{Conclusion}
We have studied the sensitivity of the LHeC to LFV dimension-six operators involving electrons and muons. Both a simple cut-based analysis and a more refined optimal observable method were employed. The results indicate that, with $3\ \text{ab}^{-1}$ of data, the LHeC can probe Wilson coefficients $C_i/\Lambda^2$ down to $\mathcal{O}(10^{-2} - 10^{-1})\ \text{TeV}^{-2}$, depending on the operator structure. Initial electron polarisation can further enhance these sensitivities, particularly for chiral operators. This complements the LFV searches at high-energy hadron and lepton colliders, and underscores the role of $ep$ machines as powerful probes of flavour-violating new physics.

\section*{Acknowledgements}
H.~Bharadwaj acknowledges the financial support from the ANRF, Govt. of India for International Travel Support (Young Scientist) ITS/2025/000234. S.~Dutta acknowledges the partial financial support from ANRF/CRG/2023/008234, Govt. of India.

\end{document}